\documentstyle[sprocl2,epsfig]{article}
\def\be{\begin{equation}}
\def\ee{\end{equation}}
\def\bea{\begin{eqnarray}}
\def\eea{\end{eqnarray}}
\begin{document}

\title{Dilute monopole gas, magnetic screening and k-tensions in hot gluodynamics}

\author{C. P. Korthals Altes}

\address{Centre Physique Th\'eorique au CNRS, Case 907, Luminy, F13288, Marseille, France\\ E-mail:altes@cpt.univ-mrs.fr}  


\maketitle

\abstracts {A dilute monopole gas explains, in quarkless gluodynamics, the small ratio $\delta$ between
the square of magnetic screening mass $m_M$ and  spatial Wilson loop tension.  This ratio is  $0.0895$ for $T=0$ to $0.0594$ at $T=\infty$ for any number of colours with order $N^{-2}$ corrections and equals up to a numerical factor of O(1) the diluteness. The monopoles have a size $l_M=m_M^{-1}$. The GNO classification tells us they are in a representation of the magnetic SU(N) group. Choosing the adjoint for the  dilute gas predicts the k-tensions to scale as $k(N-k)$ , within a percent for high T, and a few 
percent for low T for the seven ratios determined by lattice simulation. The transition is determined by the transition of the dilute Bose gas at $T_c=0.174 m_M$, and the transition is that of a non- or nearly-relativistic Bose gas.}
\section{Introduction}\label{intro}
An ancient idea in QCD~\cite{tH76} is that monopoles are responsible for flux tube formation through a dual superconductor mechanism. Another mechanism, that of the ``Copenhagen vacuum'' of the early eighties~~\cite{greensite}, proposes
macroscopic Z(N) Dirac strings or Z(N) vortices.

In this talk I will discuss a specific model
of the first type~\cite{giovanna}, that works surprisingly well in its most direct applications. A  straightforward way to see its workings is to start from the plasma phase. That sounds at first sight self-defeating as there
is no confinement in this phase. 

But it will turn out that precisely the absence of confinement  renders the detection of these monopoles, or perhaps more appropriately, magnetic quasi-particles, so straightforward.  We know that electric quasi-particles, the gluons, are approximately free at very  high temperature. This follows from the Stephan-Boltzmann law, which is verified up to $15\%$ on the lattice, at $T\sim 4T_c$. Note, though, that the interactions at this temperature are still quite strong.

 We may guess that the same happens to the magnetic monopoles that were condensed in the cold phase.  

What kind of monopoles can we expect? In the absence of any spontaneous breaking the GNO classification~\cite{olive} applies.  For a theory with only Z(N) neutral fields, like the adjoint,  the magnetic group is $SU(N)$.  So we can have monopoles in the fundamental,  adjoint or any representation 
 we like~\footnote{In the GNO analysis the range of the monopoles is infinite. That is why quarks imply a smaller magnetic group, $SU(N)/Z(N)$, because of the Dirac consistency relation between electric and magnetic charges. Also the existence of the monopoles was not proven. See e.g. ~\cite{lavelle}.}. 

One of the consequences of the idea of a monopole condensate is the screening 
of the magnetic Coulomb force between two static magnetic sources. This is a straightforward generalization of the screening of the Coulomb force between two heavy quarks. There is now ample affirmation of screening~\cite{rebbi}~\cite{deforcrand}~\cite{owe} from lattice simulations.  The magnetic screening mass $m_M$ starts out at $T=0$ and equals there  the lowest $0^{++}$ glueball mass. It stays constant  till about $2T_c$, and then starts to scale with the temperature like $g^2T$.

It is reasonable to think of the monopole as having a size of about the screening length $l_M=m_M^{-1}$. We will assume that its size is much smaller than the inter-monopole distance. That, together with the choice of multiplet, will fix the ratios of spatial string tensions. The reader who is only interested in how one computes these ratio's should read sections \ref{sec:elflux},\ref{sec:wilsonloopvalues} and \ref{sec:lattice}.

The remaining sections give a qualitative idea  of how flux loop averages
depend on the representation. The main conclusion is that the loops depend only on the N-allity. Moreover, in our model only the fully anti-symmetric representations are a practical
tool to determine numerically the tension.

\section{Electric flux loops}\label{sec:elflux}

In this section we discuss the behaviour of electric flux loops in the plasma,
 as it will explain the basic features that are permeating the talk. 
\subsection{QED}

Consider a plasma of ions and electrons. We will take the ions to have the same but opposite charge of the electron.

Suppose we want to compute the
electric flux going through some large (with respect to the atomic size) closed loop $L$
with area $A(L)$. Normalize the flux ${\mit\Phi}=\int_{\rm S} d\vec S\cdot\vec E$ by the electron charge
$e$ and define :
\begin{equation}
V(L)=\exp{i2\pi{\mit\Phi}/e}\,.
\label{classicalfluxloop}
\end{equation}

  Of course, at $T$ below the ionization temperature no flux would be detected by the loop,
because there are only neutral atoms moving through the loop. 

Let us now raise the temperature above $T_{\rm ionisation}$. What will happen?
Both electrons and ions are screened. For simplicity we will take the ions to have the
opposite
of one electron charge.

We are going to make the following simplification. The charged particles are supposed to shine their flux through the loop if they are within distance $l_D$
from the minimal area of the loop. This defines a slab of thickness $2l_D$. 

Of course if we plot $|{\mit\Phi}/e|$ as function of
the distance of the particle to the loop you find an exponential curve with the maximum
$1/2$ at zero distance. For the sake of the argument we will replace that curve by a theta
function of height $1/2$ and width $2l_{\rm D}$. If one wants to do better one has to
deal with infinitesimally thin slabs, and integrate over the thickness. This yields an effective thickness of 1.64282...~\cite{zakopane}.
Here we will keep the factor 2.
The result is parametrically the same as the one we will derive keeping the simple minded method.

Then one electron (ion) on the down side of the loop 
will
contribute $+1/2 (-1/2)$ to the flux, and with opposite sign if on the up side of the
loop. That is: $V(L)|_{\rm one\;charge}=-1$. This result is independent of the sign of the charge! The plasma is overall neutral, the loop is sensitive only to charge fluctuations. For $l$ charges inside the slab the flux adds linearly and  we find:
\begin{equation}
V(L)|_l=(-1)^l.
\label{lcharges}
\end{equation}

Assuming that all charges move independently,  the average of the flux
loop $V(L)$ is determined by the probability $P(l)$ that $l$ electrons (ions)
are present in the slab of thickness $2l_{\rm D}$ around the area spanned by the loop.  Taking
for $P(l)$ the Poisson distribution ${1\over {l!}}(\bar l)^l\exp{-\bar l}$-- $\bar l$ is
the average number of electrons (ions) in the slab--  we find for the thermal average of the
loop due to both ions and electrons:
\begin{equation}
\langle V(L)\rangle_T=\sum_lP(l)V(L)|_l=\sum_l P(l)(-)^l=\exp{-4\bar
l}\,.
\label{electrictension}
\end{equation}

Now $\bar l=A(L)2l_{\rm D}n(T)$, so the electric flux loop obeys an area law
 $\exp{-\rho(T)A(L)}$, with a tension 
\begin{equation}
\rho(T)=8l_{\rm D}n(T).
\label{classicaltension}
\end{equation}

This area law distinguishes the behaviour of the loop in the plasma 
from that in the normal unionized state.

The flux loop has a very useful alternative formulation. Introduce the charge  density $j_0$. Then you can write flux loop
as a gauge transformation $\exp{\big(-{i\over e}\int d\vec x(\vec E.\vec\nabla+j_0)\omega_1\big)}$. $\omega_1$ a gauge function that falls off fast atspatial infinity. But it has a discontinuity  $2\pi$, when crossing the surface. So the border of the surface is of course just a closed Dirac string.

If $\omega_1$ has no discontinuity, integration by parts will give us the Gauss operator $-\vec\nabla.\vec E+j_0$ in the exponent.  But the discontinuity generates the extra surface term ${2\pi\over e}\Phi$. Because the flux $\Phi$ is gauge invariant the two terms commute. So the gauge transform factorizes in a factor containing the flux and a factor containing only the Gauss operator. The latter becomes unity on the physical subspace and only the flux operator stays~\cite{kovnerrosen}.

\subsection{Electric flux in gluodynamics}\label{subsec:elfluxglue}

Is there an $SU(N)$ generalization of the QED case?
In fact yes. The closed Dirac string in QED is replaced by
a Z(N) vortex of strength $k$, the 't Hooft loop\cite{thooft}.
We introduce in the Lie algebra of traceless hermitean $N\times N$ matrices a basis for the Cartan subalgebra, the $(N-1)\times(N-1)$ dimensional subspace of diagonal matrices. This basis of $N-1$ diagonal matrices $Y_k$ is chosen such that in exponentiated form $\exp{\big(-i2\pi Y_k\big)}$ it gives the $N-1$ centergroup elements $\exp{ik{2\pi\over N}}$. A simple choice is 
\be
Y_k={1\over N}diag(N-k,\ldots{....},N-k,-k,\ldots{....},-k).
\label{ymatrix}
\ee   

The entry $N-k$ comes in $k$ times, and $-k$ comes in $N-k$ times, so the trace is $0$.

The flux operator becomes in this notation:
\be
V_k(L)=\exp{i{4\pi\over g}\int d\vec S.Tr\vec E Y_k}.
\label{elflux}
\ee

It does correspond to the vortex operator of 't Hooft with strength $k$~\footnote{Normalization is such that $[E_j^a(\vec 0),A_k^b(\vec x)]=\delta^{a,b}\delta_{j,k}\delta(\vec x)$, $\vec E=\vec E^a\lambda_a, [\lambda_a,\lambda_b]=if_{a,b,c}\lambda_c, f_{a,b,c}f_{d,b,c}=N\delta_{a,d}$.}.

The next question is: does the gas of deconfined gluons induce an area law
in this operator? The answer is yes, and the reasoning is as before.
The charge of a gluon with respect to the charge $Y_k$ is found from the
form $Y_k$ takes in the adjoint representation. This is easy: we have charge $\pm 1$ like in QED, but unlike in QED we have $2k(N-k)$ gluon species
in the gluon multiplet with such a charge. All other gluons have vanishing charge so do not contribute to the flux. Since we take the gluons to be statistically independent, the charged ones all contribute a factor to $\langle V_k\rangle$, and this factor is the same by $SU(N)$ symmetry: $\exp{-2l_Dn}$.

Conclusion: the expectation value of the loop is
\be
\langle V_k\rangle=\exp{\big(-\rho_k(T)A(L)\big)} 
\ee
\noindent with the tension 
\be
\rho_k=4l_Dnk(N-k).
\ee

So this is the k-scaling law for the electric flux loop. It does obey large $N$ factorization:
\be
 \rho_k=k\rho_1.
\label{factorization}
\ee

 But it has $1/N$ corrections and not, as perhaps expected  in gluodynamics, $1/N^2$. 

This computation is corroborated by low order perturbation theory for the loop. The expectation value of the loop can be computed from a tunneling process between adjacent vacua of an effective potential with a $Z(N)$ symmetry~\cite{bhatta},~\cite{korthalskovner}. This potential has, because of the $Z(N)$ symmetry, degenerate minima in the centergroup elements. Only at  cubic order in the coupling $g$ the k-scaling starts to deviate slightly ~\cite{giovanna}. Simulations in $SU(4)$  show k-scaling up to $2T_c$ ~\cite{biagio}.

The effective potential is computed in a background that serves as profile orthogonal to the area. The minimal effective action of this profile gives the tension $\rho_k$, thereby tunneling from say $1$ to $\exp{i2\pi Y_k}$. The profile is present in the propagators and renders the usual double line
representation of the gluon propagators invalid, hence the $1/N$ corrections.
The typical width of the profile around the area is the Debye screening length $l_D$.

\subsection{Junctions and highest weights}\label{subsec:junction}

 It may have struck the reader that we used $Y_k$, instead of 
$kY_1$ in the flux formula eq.(\ref{elflux}).  The reason is that the tunneling process from $0$ to $Y_k$ passes only one mountain. On the other hand tunneling from $0$ to $kY_1$ passes through k mountains, each of which interposes between two subsequent $Z(N)$ vacua and contributes 
$\rho_1$.  So at first sight we find $k\rho_1$ from this operator. However,
there must be a {\it{ lower}} state $\rho_k$, as the following argument shows.
Formally the operator 
\be
V_k(L)=\exp{i{4\pi\over g}\int d\vec S.Tr\vec E k Y_1}
\label{kelflux}
\ee

\noindent is obtained by coalescing the product of $k$ operators $V_1(L)$.
There are two relevant stages here: one where all the loops are still more than the screening distance away from each other. Then every loop will form its individual
 profile and wall, and give a factor $\exp{-\rho_1A(L)}$. The second stage is when all
 loops are within screening distance. Then the individual walls emanating from each loop will form a junction at a typical distance $l_D$. At this junction
 k unit walls merge with the k-wall, and the k-wall views the ensemble of the
k unit discontinuities as one discontinuity of strength k, on the scale of the screening length. The junction coupling involved here is non-perturbative. It may be small, certainly in our model of approximately free gluons. Hence the true $\rho_k$ may be difficult to disentangle numerically.

In ref.~\cite{kovnerfossil} the case $k=N$ in eq.(\ref{kelflux}) is discussed. The operator $V_{k=N}$  creates a state with energy $N\rho_1$ and involves the special junction where $k=N$.
This junction has only N incoming k=1 walls, and no outgoing N wall. That explains the periodicity  $N$ of the tension $\rho_k$. 

In addition we can imagine junctions where two of the incoming unit walls are replaced by one k=2 wall. That junction  corresponds to the matrix $(k-2)Y_1+Y_2$.

A general junction in SU(N) will be specified by a set of non-negative numbers $w_1,w_2,.....,w_{N-1}$. So the constraint on those numbers is:
\be
 \sum_{l=1}^{N-1}lw_l=k.
\ee

\begin{figure}
\begin{center}
\epsfig{figure=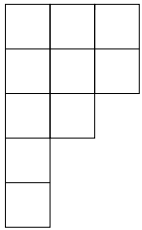, height=3cm} 
\caption{A Young tableau with $w_1=0,w_2=1=w_3,w_4=0=w_5,w_6=1, H_R=Y_2+Y_3+Y_6$ and $k=10$.}
\label{fig:young}
\end{center}
\end{figure}

In our two previous examples we had $w_1=k$ all others zero, and $w_1=k-1, w_2=1$ , all others zero.
The corresponding matrix is then $\sum_{l=1}^{N-1}w_lY_l$ and the reader will recognize this matrix as being the highest weight $H_R$ of the Young tableau characterized (see fig.\ref{fig:young})  by the numbers $w_1,.....w_{N-1}$!
These Young tableaux refer to the representations of the magnetic GNO group.

 Also  this general junction $J(k;\{w\})$ is  expected  to be small for the same reasons. Especially, because of factorization, eq. (\ref{factorization}), it vanishes in the large N limit. The conclusion is that only for the partitioning $w_k=1$ , all other $w$'s zero, we obtain a direct coupling to the k-tension, and no junction is needed. All other charges
 will give states of non-interacting k-tensions. 

Something very similar happens for the various representations involved in the 
Wilson loops and we will come back to this in detail in section~\ref{subsec:nallity}.

\section{Some basic facts in quarkless Yang-Mills}\label{basic}
In this section we gather a few facts, partly empirical, from lattice 
simulations, and partly from simple physics arguments.
\subsection{Magnetic screening}\label{subsec:screening}
A feature that sets gluodynamics apart from QED is the appearance
of magnetic screening. When we put two heavy monopoles far apart at 
a distance $r$ we find a Yukawa type potential:
\be
V(r)={c\over r}\exp{-m_Mr}.
\label{magneticyukawa}
\ee
 A simple argument shows~\cite{zakopane}~\cite{owe}  that the magnetic mass is the mass  of a scalar
excitation of a Hamiltonian in a world with two large space dimensions and
one periodic mod $1/T$. The space symmetries in such a world are $SO(2)\times
 P\times C\times R$. We have the rotation group in the two large dimensions, C is charge conjugation, P is parity in the large 2d space, and R is the parity in 
the periodic direction and a state is written as $J^{PC}_{R}$~\cite{arnold}. The state corresponding to the magnetic mass is $0^{++}_{+}$. As $T\rightarrow 0$
 the rotation group becomes $SO(3)$, and $R$ and $P$ become related through a rotation, so at $T=0$ the $0^{++}$ glueball mass results.     The 
Debye mass, screening the electric Coulomb force between two heavy quarks, is related to the 
$0^{+-}_{-}$ state, and disappears below $T_c$. Below $T_c$ $Z(N)$ symmetry is restored and the string tension appears, as a state transforming non-trivially
under $Z(N)$.  

In contrast, the magnetic mass does not feel the $Z(N)$ transition. At  temperatures where $T$ dominates all scales, the magnetic mass starts to scale like $g^2T$,
as schematically shown in fig.~\ref{fig:fig1}.

\subsection{Spatial  tension}\label{subsec:sigmas}

The spatial  tension is defined through a spatial loop $L$, on which an
ordered product ${\cal{P}}\exp{i\oint d\vec s.\vec A_R}$ lives. $R$ labels the representation for the vector potential we have chosen. The spatial Wilson loop is then defined as the trace over colour:
\be
W_R(L)=Tr{\cal{P}}\exp{i\oint_L d\vec s.\vec A_R}.
\label{wilsonordered}
\ee

\begin{figure}[htb]
\begin{center}
\epsfig{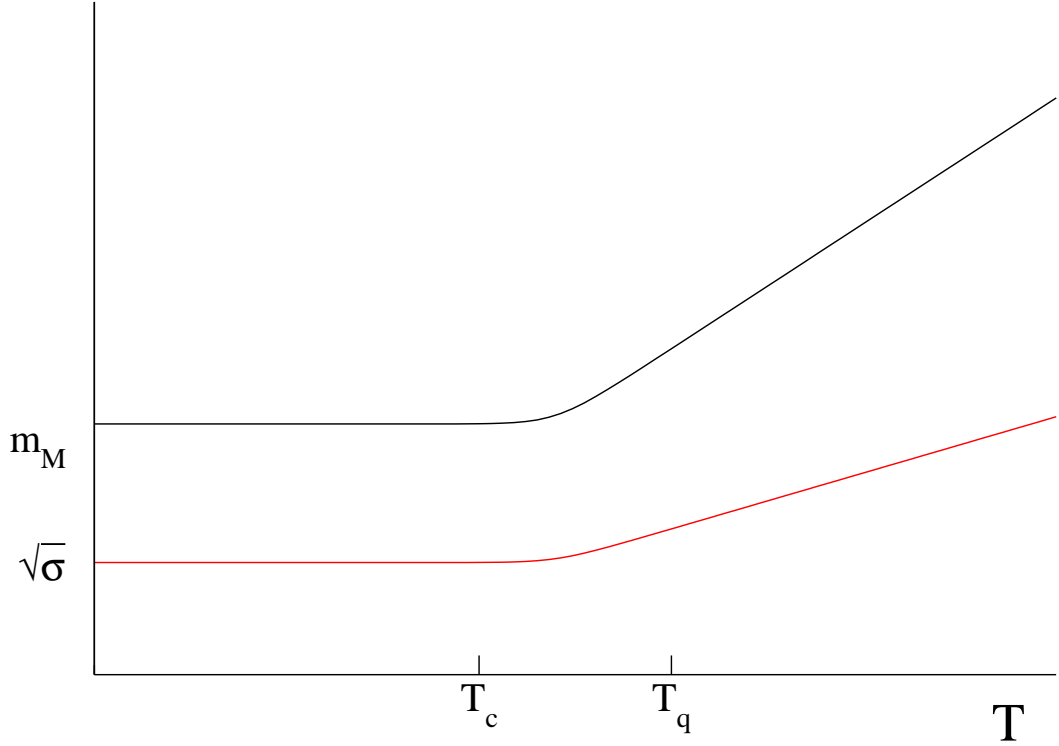}
\caption{Magnetic mass $m_M$ and tension $\sigma$ as function of temperature, schematically. $m_M(0)=m_{0^{++}}$ and from ref. (13): $T_c=0.174~m_{0^{++}}$.
 The temperature $T_q\le m_{0^{++}}$ is where the de Broglie thermal wave length becomes equal to the magnetic screening length. For the calculation of the  tension it is  below $T_q$ that quantum statistics applies, above classical statistics applies as in section 5.}
\label{fig:fig1}
\end{center}
\end{figure}

The path integral average gives an area  law:
\be
\langle W_R(L)\rangle=C\exp{\big(-\sigma_R(T) A(L)\big)}
\label{wilsonarealaw}
\ee  

\noindent for all temperatures. At zero temperature the spatial  tension
equals the string tension~\footnote{In what follows we reserve the name string tension for the time-space loop.} from a time-space loop because of Euclidean rotation invariance. But as function of temperature they behave differently. The string tension suffers a correction at low temperature of $-{\pi\over 3}T^2$ due to the excitations contributing the Luescher term. Above $T_c$ it vanishes. The spatial tension stays flat, see fig. (\ref{fig:fig1}). Like the magnetic mass it does not feel the $Z(N)$ transition and  starts to scale like $(g^2T)^2$ well above the critical temperature.
The dimensionless ratio of the spatial tension in the fundamental representation and the square of the magnetic mass is small, on the order of a few percent.
In two extreme cases the ratio $\delta(T)$ is  accurately known from simulations : at $T=\infty$~\cite{teper3d}~\cite{owe} in the 3d theory,  and at
$T=0$~\cite{teper4d}. In between there are simulations~\cite{rebbi}~\cite{deforcrand}
that are consistent with a slowly varying $\delta(T)$. It is very important to have this verified with more precision.

\subsection{Dependence of the tension on the representation, and N-allity}\label{subsec:nallity}

There is widespread agreement on the dependence of the tension and the string tension on the representation $R$. It  is only its N-allity $k$ that counts. If we think of R being built up by $f$ fundamental and $\bar f$ anti-fundamental representations, the N-allity is just the difference:
\be
k=f-\bar f.
\ee 

To get a more precise idea, let us imagine the Wilson loop is formulated 
in a periodic box, with the 3 space dimensions being of macroscopic length
$L$. The fourth dimension is of length $1/T$. The loop is now replaced by
two Polyakov lines of opposite orientation in the x-direction. They are a distance r apart- and carry the reprsentation $R$. The expectation value of the loop is then parametrized as:
\be
\langle W_R(L)\rangle=C\exp{-LV_T(r)}.
\label{pot}
\ee

If the distance $r$ is very small, asymptotic freedom tells us:
\be
V(r)={g^2(r)C_2(R)\over {4\pi r}}
\label{shortdistancepot}
\ee
\noindent where $C_2(R)$ is the quadratic Casimir operator.

In three dimensions the Coulomb force is logarithmic.

The Casimir  operator can be related to our $N\times N$ matrices $Y_k$
if  $R$ corresponds to a Young tableau, with the first row
having $w_1$ more boxes than the second. The second row has $w_2$ more
boxes than the third, and so on. By definition, the numbers $w_k$ are never negative.

There is a one-to-one correspondence between Young diagrams and irreducible
representations of SU($N$).

Define the highest weight $H_R$ of $R$
as 
\be
H_R=\sum_kw_kY_k 
\label{highweight}
\ee

\noindent as we already alluded to at the end of section~\ref{subsec:elfluxglue} . An example is shown in fig.\ref{fig:young}.

Then, if $Y=\sum_kY_k$ :
\be 
C_2(R)={1\over 2}(TrH_R^2+2TrYH_R).
\label{secondcasimir}
\ee

If the distance is long enough ( a somewhat ambiguous criterion!) the string regime sets in and we expect:
\be
V(r)=\sigma_k(T)r.
\label{ktension}
\ee

The tension $\sigma_k$ only depends on the N-allity $k$. String formation between the two sources renders this dependence very plausible. The 
tension $\sigma_1$ of the string formed in between two fundamental sources can form a bound state in the string with tension $\sigma_2$. We can go on like this and the question is, what is the dependence on $k$? It should obey certain a priori criteria. For large number of colours we expect factorization
\be
\sigma_k=k\sigma_1.
\label{factorize}
\ee

From SUSY experience one also expects~\cite{shifman} in that same limit:
\be
\sigma_k=k\sigma_1+O(1/N^2).  
\label{1overn}
\ee
 This is discussed in depth  by Misha Shifman in this volume. 

\subsection{Flux representation of the Wilson loop}\label{fluxdiakonov}

If we want to proceed with the Wilson loops, as we did with the 't Hooft loops,
we need clearly a representation similar to eq.(\ref{elflux}).

Remember we discussed in section \ref{subsec:junction} a general formula for the electric flux loop with $Y_k$ in eq.(\ref{elflux}) replaced  by $H_R$,the highest weight but still with N-allity $k$:$V_R\exp{({4\pi\over g}\int d\vec S.Tr\vec E H_R)}$.

Of course a natural guess is to take this electric flux formula, replace $\vec E$ by $\vec B$, $\alpha=g^2/4\pi$ by $1/\alpha$ to get:
\be
W_R~''=''\exp{(ig\int d\vec S.Tr\vec B H_R)}.
\label{simpledual}
\ee

This formula cannot be true~\cite{korthalskovner}. But it is true in a dynamical sense: consider the limit of a   path integral average in the 3d theory with an adjoint Higgs, the electrostatic QCD Lagrangian.

  For a given representation $R$ with highest weight $H_R$ with stability group $S$ (i.e. $SH_RS^{-1}=H_R$) one finds:

\be
\langle W_R\rangle=<\int D\Omega \exp{\big(ig\int d\vec S.(\vec B_a-{1\over g}f^{abc}\vec D n_b\wedge\vec D n_c)Tr{\lambda_a\over 2}\Omega H_R\Omega^{\dagger}\big)}>. 
\label{magflux}
\ee

 \noindent with $n_a=Tr{\lambda_a\over 2}\Omega H_R\Omega^{\dagger}$  the Higgs field's angular part that parametrizes the coset $SU(N)/S(H_R)$.
The second term in the exponent is the source term for the monopoles in the sense that it carries no long range effects in the original Higgs phase. The average of the r.h.s is the 3d Yang-Mills average. The integration is over all gauge transforms with an arbitrary number of hedge hog configurations. The r.h.s. is obtained ~\cite{korthalskovner2} by taking the average of the magnetic charge operator in the 3d Higgs phase characterized by $H_R$. Then the the VEV of the Higgs field is let to zero, which introduces the fluctuations over $n_a$, the angular components of the Higgs field.  Finally one lets the mass of the Higgs become infinite, which suppresses the radial integrations.

On the other hand~\cite{diakonov} the integrand of the r.h.s., without the brackets, was shown by Diakonov and Petrov to equal $W_R$, by one dimensional quantum mechanics methods. It involves a limiting procedure which is procured in the path integral by  starting out from the Higgs phase and moving to the symmetic phase as described above. With this specification we will use eq. (\ref{simpledual}) without quotation marks in what follows.

\begin{figure}
\begin{center}
\epsfig{figure=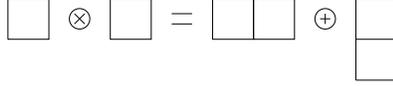, angle=270, width=5.2cm} 
\caption{Young tableaux with k=2}
\label{fig:fig2}
\end{center}
\end{figure}

So we need a stable Higgs phase for the limiting procedure.
The question is then: what are the stable Higgs phases? The answer is from simulations:
only those that have a stability group $S=SU(l)\times SU(N-l)\times U(1)$~\cite{rajantie}.

Summarizing: it is possible to have a magnetic flux representation  
in a path integral average for Wilsonloops $W_R$ with
special highest weight $H_R$. The highest weight should commute with 
a subgroup of the form $SU(l)\times SU(N-l)\times U(1)$. These invariance groups define the only stable Higgs phases in 3d.  The monopoles in these phases 
are partly screened in the strongly coupled sectors ( the non-abelian factors 
in the stability group). The  SU(5) case is nicely described by Coleman in ref.\cite{olive}. After the limiting procedure they are  the fully screened monopoles in the symmetric phase.

To get a feeling, look at the k=2 representations of $SU(N)$ in fig.~\ref{fig:fig2}).
The highest weight of the totally anti-symmetric representation is computed from eq.(\ref{ymatrix}) and eq.(\ref{highweight}). It is $Y_2$, and $SU(2)\times SU(N-2)\times U(1)$ is its invariance group.  So it has a flux representation
$W_R=\exp{ig\int d\vec S. Tr\vec B Y_2}$.

The symmetric k=2 representation has highest weight $2Y_1$. It has invariance group $SU(N-1)\times U(1)$. So its flux representation is 
 $W_R=\exp{ig\int d\vec S. Tr\vec B 2Y_1}$.

\section{Monopoles and the magnetic group}\label{sec:magneticgroup}

In this section we briefly touch on  a medley of problems related to magnetic
screening.

The generators of the magnetic group as advocated by Goddard et al.~\cite{olive} are 
not known in terms of the Yang-Mills potentials. In our opinion the monopoles
are collective excitations of the magnetic gluons and can be much better 
called magnetic quasi-particles. They are bound states of magnetic gluons. Their  size is supposed to be $l_M$, the magnetic screening length. 

The question is then: what representation of the magnetic group is realized 
by Nature for these dynamical quasi-particles? 
 We have tried in the next section the simplest ones, the fundamental and the adjoint.
The adjoint has the advantage that it is compatible with quarks. Compatible
means the Dirac consistency condition is fulfilled. As already mentioned
in  the introduction, the consistency condition is stricly enforced 
in the absence of screening. If screening is present a famous example due to
't Hooft~\cite{thooftmagn} tells us that screening renders the Dirac condition
{\it less} restrictive than naively thought. At any rate the adjoint is clearly favoured by simulations of the Wilson loops, see section~\ref{sec:lattice}.

Amusing, though perhaps academic, is the observation that the flux representation of the 't Hooft loops introduced highest weights $H_R$. These should correspond to representations of the magnetic group. If we could implement the loop
with a given weight $R$ by means of a magnetic gauge potential we would have an alternative expression for the 't Hooft loop. Once the magnetic group becomes a gauge group with gauge potentials one has an alternative for the Wilson loops as well, to wit, as $Z(N)$ discontinuous gauge transformations like the 't Hooft loops.  The magnetic gauge group provides long range excitations in the low $T$ phase in contradiction with the full magnetic screening. They should be obliterated by a Higgs mechanism that breaks fully
 the symmetry. And for that our adjoint multiplet is not a good candidate. It will always leave some some subgroup unbroken as we discussed in the previous section.  A fundamental multiplet would fully screen, but is excluded by the lattice data.

\section{Predictions for the Wilson loops}\label{sec:wilsonloopvalues} 

Once we are given a dilute gas of screened monopoles, we only need to specify
the representation of the magnetic group for our monopoles. Then the 
calculation of the tension is  done with our flux representation
for the average of the loop, eq.(\ref{magflux}). But in practice we can use 
the simple formula eq.(\ref{simpledual}). The reason is that we put the contribution of the monopole gas in by hand. That takes care of the singular gauge transforms in eq.(\ref{magflux}). The remaining regular ones can be dropped
because we compute something gauge invariant.

So the average of a k-loop in the totally anti-symmetric representation
is computed from
\be
\langle W_k\rangle=\langle \exp{ig\int d\vec S.Tr\vec B Y_k}\rangle.
\label{askav}
\ee

For  the dilute gas in the adjoint representation the computation is not 
different
from the one with gluons in section \ref{subsec:elfluxglue}. The adjoint monopoles 
have 
a magnetic charge equal to $\pm{2\pi\over g}$, or $0$. Only the former contribute and their individual  contribution to the loop in eq. (\ref{askav}) is $-1$. The diluteness, and classical
statistics (as the thermal wave length is $T^{-1}$ and the typical interparticle distance is $(g^2T)^{-1}$ this is justified, see also fig.~\ref{fig:fig1}.) then give for every charged species the same contribution $\exp{-2l_Mn_M}$, $n_M$ being the common density of a given species. As there are $2k(N-k)$ charged species in the adjoint,
 the total results in the k-tension being:
\be
\sigma_k=4k(N-k)l_Mn_M.
\label{adjointtension}
\ee 

One can do a similar calculation for monopoles in the fundamental
 representation. The counting goes now as follows. Recall eq.(\ref{ymatrix}),
$Y_k={1\over N}diag(N-k,....N-k,-k,....-k)$ with trace zero.  The $Y_k$ charge
 of the highest $k$ components of
the $N$ components of the column spinor is given by ${(N-k)\over N}$. The remaining $N-k$ components have charge $-{ k\over N}$.

 It then follows from the use of the Poisson distribution that
the flux of a given component is contributing $\cos(\pi {(N-k)\over N})$ 
or $\cos(\pi{ k\over N})$. Taking into account the degeneracies the final result for the tension becomes:
\be
\sigma_k=4l_Mn_M(N-k\cos(\pi {(N-k)\over N})-(N-k)\cos(\pi{ k\over N})).
\label{fundamentaltension}
\ee

Both tensions give ratios $\sigma_k/\sigma_1$ that  behave like $k$ for large $N$ and finite $k$. This  factorization is  expected for a k-loop tension caused by screened particles~\footnote{Factorization is not obvious if uncorrelated 
Z(N) vortices~\cite{greensite} cause the area law: $\sigma_k\sim (1-cos(2k\pi/N))$. This is due to the macroscopic size of the vortex perimeter causing long range correlations between loops.} . But for large $N$ and $k=N/2$  the ratio for the fundamental multiplet is a factor 2 larger.

\section{Comparison to lattice simulations}\label{sec:lattice}

We have been discussing a model at very high temperature. Hence it is tested 
in 3d lattice simulations.
The ratios found~\cite{teper} for the totally antisymmetric irreps are close --- within a percent
for the central value --- as far as the adjoint multiplet of magnetic quasi-particles is concerned :
\begin{eqnarray*}
{\rm SU}(4):\sigma_2/\sigma_1&=&1.3548\pm 0.0064~~~~\hbox{adjoint} :1.3333~~~~~\hbox{fundamental} :1.8182 \\
{\rm SU}(6):\sigma_2/\sigma_1&=&1.6160\pm 0.0086~~~~\hbox{adjoint} :1.6000~~~~~\hbox{fundamental} :1.9686 \\
            \sigma_3/\sigma_1&=&1.808~\pm 0.025~~~~~~\hbox{adjoint} :1.8000~~~~~\hbox{fundamental} :2.3635
\end{eqnarray*}

The results are that precise, that you see a two standard deviation from the adjoint, except
 for the second
 ratio of $SU(6)$. This deviation is natural, since the diluteness of the
magnetic quasi-particles is small, on the order of a couple of percent, as we will explain at the end of this subsection. So we expect corrections on that order to our ratios. 

There is a less precise determination of the ratio
 $\sigma_2/\sigma_1=1.52\pm 0.15$ in
$SU(5)$~\cite{meyer}. But the central value is within $1$ to $2\%$ of the
predicted value $3/2$ from the adjoint. The fundamental gives a ratio $1.8231$.

The $SU(8)$ ratios are known on a rather course lattice~\cite{meyer} and using a different algorithm:

\begin{eqnarray*}
\sigma_2/\sigma_1 & = & 1.692(29)~~~~\hbox{adjoint} : 1.714~~~~~\hbox{fundamental} : 2.106\\
\sigma_3/\sigma_1 & = & 2.160(64)~~~~\hbox{adjoint} : 2.143~~~~~\hbox{fundamental} : 2.958\\
\sigma_4/\sigma_1 & = & 2.26(12)~~~~~\hbox{adjoint} : 2.286~~~~~\hbox{fundamental} : 3.256\\
\end{eqnarray*}

In conclusion: the seven measured  ratios are consistent with the quasi-particles being
independent, as in a dilute gas and in the adjoint representation. The number of quasi-particle species contributing
to the k-tension is $2k(N-k)$. This number happens to coincide with the quadratic Casimir operator of the anti-symmetric representation.

The fundamental monopoles are clearly disfavoured by the data.

For other representations we have already said what to expect. E.g. the fully symmetric representation with k boxes  will show the same tension, but through a very suppressed junction, especially for large N~\footnote{For SU(3) the k=2 symmetric representation still has an appreciable junction value, see ref.~\cite{deldebbiosu3}}. So lattice data will show predominantly the $k(\sigma_1)$  tension. 

A caveat is in order. As we are in three dimensions, the Coulomb law is logarithmic. It is multiplied by the quadratic Casimir $C_2(R)$.  So logarithmic confinement follows Casimir scaling!

So if the fully symmetric representation with k boxes
shows ratios compatible with the $C_2(R_{sym})\sim k(N+k)$ the strings are probably behaving somewhere in between logarithmic and linear confinement. 

 An inveterate pessimist could say the same of the antisymmetric reps where the Casimir ratio happens to be the same as for our adjoint monopole model. According to him, in the fitting  of the tension the plateau has been mistaken for a linear law, whereas in reality it is logarithmic. I do not know whether the data exclude this. 

\section{Conclusions}

 There is remarkable  agreement with an accuracy of about a percent between numerical simulation at high $T$ and the dilute adjoint monopole gas. This diluteness appears as a small parameter for {\it every} SU(N) theory. Some parameters, like the mass of the magnetic quasi-particle, are still within a large range, although its size is $l_M$. Is it heavy, on the order of the lowest glueball mass (i.e. $m_M$), then the transition is that of a non-relativistic Bose gas. Is it light (with respect to its size)
then the transition is that of a near-relativistic BE transition.

Perhaps the best way to summarize our approach is to return to fig.~\ref{fig:fig1}. It is clear that the 
relation:
\be
\sigma\sim l_Mn_M
\ee
\noindent implies that $l_M^2\sigma$ is the diluteness $\delta$, and must be small for consistency. And lattice data have borne that out! Also flux tube models
like that of Isgur-Paton~\cite{paton} predict this small ratio~\cite{teperjohnson} as a result of the balance of the string force and the phonon excitations of the closed string forming the lowest glueball.

Now , once we accept the idea, that on the high temperature side of the 
transition a dilute gas describes things so well, the constancy of that diluteness down till $T=0$ suggests that all what changes is the thermal wavelength
$\lambda_B(T)$. At temperatures on the order of the glueball mass or higher
it is clear that the interparticle distance $O(1/g^2T)$ is  larger than the thermal 
wave length $1/T$, because the coupling is so small. But at a temperature $T_q$
the thermal wave length takes over with $T_q\sim m_{monopole}$. It is very unlikely that $T_q$ will be below $T_c$, since that would imply the BE transition as
a second one below $T_c$. So the monopole mass should not be lower than $T_c$,    
so the transition will then be a non- or near-relativistic BE transition. 

For the convenience of the reader we give the lattice numbers:\linebreak

 $T_c=\big(0.596(4)+0.453(30)/N^2\big)\sigma_1$~\cite{teperT}, and 
$m_{0^{++}}/\sigma_1= 3.341(76)+1.75/N^2$~~\cite{teper4d} or $T_c=0.174(5)(1+0.27(3)/N^2)m_{0^{++}}$. This gives the non-relativistic limit if the monopole is as heavy as $m_{monopole}=m_{0^{++}}=l_M$. In the unlikely case that  $m_{monopole}$ is as light as $T_c$ we have the near-relativistic case.

 Below $T_q$ the dilute gas gives a contribution to our ratios,
but now determined by Bose statistics. For an analysis of the cold phase ratios see ref.~\cite{korthalsmeyer}. 

Of course for most  observables in the low $T$ phase the Bose statistics is all-important. Thus the string tension will become non-zero below $T_c$, the character of the transition, i.e. a jump  or continous behaviour in the occupation fraction of the $\vec p=0$ states will be crucial to know and and is calculable in this model. Vortices in the condensed superfluid  give then the flux tubes in the Isgur-Paton model of QCD.

Realistic QCD involves quarks, and there is all reason to believe they couple strongly to our monopoles. After all, the latter are  bound states of magnetic gluons. In the CFL phase colour is fully broken~\cite{alford}, and it is amusing to speculate on how the adjoint monopoles behave there. A fundamental multiplet of monopoles would get confined.

 It would be interesting to  check this model in SUSY gluodynamics, where so many features are known
analytically.

The unbearable heaviness of the ground state (the energy density of our dilute gas is on the order of $1 (GeV)^4$) can be taken into account by
string theory calculations of the same ratios~\cite{herzog}. Indeed, taking into account gravitational effects in an AdS/CFT context, does not affect the result gotten by the simple monopole gas picture: $k(N-k)$ scaling does result in a large $N$ limit, with $k/N$ fixed. Note however that in practice these calculations are done in weak gravitational bulk fields.

Note that a real time picture of our quasi-particles is lacking. According to our Euclidean picture they become, at very high $T$, a 3d gas of particles with
small size $l_M$ and small interparticle distance. The real time picture is a challenge.

 Luigi del Debbio, Pierre Giovannangeli, Christian Hoelbling, Dima Kharzeev, Alex Kovner, Mikko Laine, Biagio Lucini, Harvey Meyer, Rob Pisarski, and Mike Teper provided me with useful comments.
  
I thank the organizers for  their invitation, an inspiring meeting, and for wonderful hospitality.


\section*{References}

\begin{thebibliography}{0}

\bibitem{tH76} G.~'t Hooft, in High Energy Physics, ed. A. Zichichi (Editrice
Compositori Bologna, 1976);
S.~Mandelstam, Phys.~Rep.~23C (1976), 245
T.~G.~Kov\'acs and E.~T.~Tomboulis, Phys.~Rev.~D57 (1998) 4054;
 T.~G.~Kov\'acs and E.~T.~Tomboulis, Phys.~Lett.~B463 (1999) 104;
J.~M.~Cornwall, Phys.~Rev.~D57 (1998) 7589;
J.~M.~Cornwall, Phys.~Rev.~D58 (1998) 1250.
J. M. Carmona, M. d'Elia, L. del Debbio,A.  di Giacomo, B. Lucini, G. Paffutti,
Phys.Rev.D66:011503,2002; hep-lat/0205025.
J. Smit, A. van der Sijs, Nucl.Phys.B355,603,(1991). 
\bibitem{greensite}
H. B. Nielsen, P. Olesen, Nucl. Phys.B61, 45,(1973); Nucl. Phys.B160, 380 (1979).
R.~Bertle, M.~Faber, J.~Greensite, S.~Olejnik,
  Nucl.~Phys.~Proc.Suppl. 83 (2000) 425;
 M.~Engelhardt, K.~Langfeld, H.~Reinhardt and O.~Tennert,
Phys.~Lett.~B431 (1998) 141.
\bibitem{thooft}G.'t Hooft,  Nucl.Phys.B138, 1 (1978).
\bibitem{giovanna} P. Giovannangeli, C.P. Korthals Altes, Nucl.Phys.B608:203-234,2001;  hep-ph/0102022.
\bibitem{olive}P. Goddard, J. Nuyts, D. Olive, {\it{Nucl.Phys}} {\bf{B125}}, 1 (1977). S.Coleman, Erice lectures 1981.
\bibitem{thooftmagn}G. 't Hooft, Nucl. Phys.B 105, 538 (1976).
\bibitem{lavelle}A. Kovner, M.Lavelle, D. McMullan, JHEP 0212:045, 2002; hep-lat/0211005.
\bibitem{owe} P. de Forcrand, C. P. Korthals Altes, O. Philipsen, in preparation\bibitem{arnold} P. Arnold, C. Zhai, Phys.Rev.D51:1906-1918,1995 
; hep-ph/9410360. 
\bibitem{zakopane}C. P. Korthals Altes, 2003 Zakopane lectures,hep-ph/0406138.
\bibitem{teper3d} M. Teper, Phys.Rev.D59, 014512, 1999; hep-lat/9804008. 
\bibitem{teper4d} B. Lucini, M. Teper, JHEP 0106,050 (2001), M. Teper, hep-th/9812187 .
 \bibitem{teperT} B. Lucini,  M. Teper, U. Wenger,  JHEP 0401:061,2004, hep-lat/0307017. 
\bibitem{teper} M. Teper, B. Lucini, Phys.Rev. D64 (2001) 105019.
\bibitem{rebbi}C. Hoelbling, C. Rebbi, V.A. Rubakov, Phys. Rev.D63 :034506,2001; hep-lat/0003010; C. Hoebling, in preparation.
\bibitem{shifman} A. Armoni, M. Shifman, 
Nucl.Phys.B671, 67, 2003; hep-th/0307020. 
\bibitem{diakonov}D. Diakonov, V. Petrov, Phys. Lett.B242, 425 (1990); see also hep-lat/0008004.
\bibitem{kovneraltes}C.P. Korthals Altes, A. Kovner, Phys.Rev.D62:096008, 2000; hep-ph/0004052 
\bibitem{biagio} B. Lucini, Ph. de Forcrand, private communication.
\bibitem{meyer}H. Meyer, private communication and hep-lat/0312034.
\bibitem{korthalsmeyer}C. P. Korthals Altes, H. Meyer, to be published.
\bibitem{bhatta} T. Bhattacharya, A. Gocksch, C. P. Korthals Altes and R. D. Pisarski, 
 Phys.Rev.Lett.66, 998 (1991);
Nucl.Phys.B 383 (1992), 497.
\bibitem{korthalskovner2} C. P. Korthals Altes, A. Kovner, Phys.Rev.D, 62,096008.\bibitem{korthalskovner}C.P. Korthals  Altes, A. Kovner and M. Stephanov, hep-ph/99, Phys.Lett.B469, 205 (1999), hep-ph/9909516. 
\bibitem{deforcrand}Ph. de Forcrand, M. D'Elia, M. Pepe, Phys.Rev.Lett.86:1438,2001; hep-lat/0007034. 
\bibitem{kovnerfossil}A similar problem is discussed by A. Kovner, Phys.Lett.B509:106-110,2001; hep-ph/0102329. 
\bibitem{kovnerrosen}For a related problem see: A. Kovner, B. Rosenstein, Int.J.Mod. Phys. A7(1992), 7419.
\bibitem{rajantie} A. Rajantie, Nucl.Phys.B501 521, 1997; hep-ph/9702255; Ling-Fong Li, Phys.Rev.D9 1723, 1974; H. Ruegg, Phys.Rev.D22 (1980), 2040;K. Olynyk, J. Shigemitsu, Phys. Rev. Lett. 54, 2403 (1985).
\bibitem{deldebbiosu3} L.del Debbio, H. Panagopoulos, E. Vicari, hep-lat/0308012.\bibitem{paton}N. Isgur, J. Paton, Phys. Rev. D31 (1985), 2910.
\bibitem{teperjohnson} R.W. Johnson, M. Teper, Nucl.Phys.Proc.Suppl.63:197,1998; hep-lat/9709083 
\bibitem{korthalsgiovanna}P. Giovannangeli, C.P. Korthals Altes, hep-ph/0212298 computes in large $N$ limit, and to appear soon.
\bibitem{herzog}Chris P. Herzog,
 Phys.Rev.D66:065009,2002; hep-th/0205064 
\bibitem{alford}M. G. Alford, K. Rajagopal, F. Wilczek, Nucl.Phys.B537:443,1999; hep-ph/9804403. 
\end{thebibliography}
\end{document}